\documentclass[useAMS,usenatbib]{mn2e}                                                             
\usepackage[totalwidth=515pt,totalheight=660pt,left=1.4cm,right=1.4cm]{geometry}              
\usepackage{graphicx,amssymb,color}                                                         
\usepackage[normalem]{ulem}                                                                    
\input psfig 

\title[Unpacking planetary systems around WDs]
{Detectable close-in planets around white dwarfs through late unpacking}
\author[Veras \& G\"{a}nsicke]{
Dimitri Veras$^{1}$\thanks{E-mail: d.veras@warwick.ac.uk},
Boris T. G\"{a}nsicke$^{1}$
\\
$^{1}$Department of Physics, University of Warwick, Coventry CV4 7AL, UK
}

\begin{document}

\date{Accepted 2014 November 21. Received 2014 November 18 in original form 2014 October 26}

\pagerange{\pageref{firstpage}--\pageref{lastpage}} \pubyear{2015} 
%\onecolumn

\maketitle

\label{firstpage}

\begin{abstract}
Although 25\%-50\% of white dwarfs (WDs) display evidence for remnant planetary systems, their orbital architectures and
overall sizes remain unknown.  Vibrant close-in ($\simeq 1 R_{\odot}$) circumstellar activity is detected at WDs spanning many
Gyrs in age, suggestive of planets further away.  Here we demonstrate how systems with 4 and 10 closely-packed 
planets that remain stable and ordered on the main sequence can become unpacked when the star evolves into a WD and 
experience {\it pervasive inward planetary incursions throughout WD cooling}. Our full-lifetime simulations run for the age 
of the Universe and adopt main sequence stellar masses of $1.5M_{\odot}$, $2.0M_{\odot} $ and $2.5M_{\odot}$, which correspond 
to the mass range occupied by the progenitors of typical present-day WDs.  These results provide (i) a natural way to 
generate an ever-changing dynamical architecture in post-main-sequence planetary systems, (ii) an avenue for planets to
achieve temporary close-in orbits that are potentially detectable by transit photometry, and (iii) a dynamical 
explanation for how residual asteroids might pollute particularly old WDs.
\end{abstract}

\begin{keywords}
minor planets, asteroids: general -- stars: white dwarfs -- methods:numerical -- 
celestial mechanics -- planet and satellites: dynamical evolution and stability
-- protoplanetary discs
\end{keywords}

\section{Introduction}

Our Solar system is packed.  In other words, it could quickly become dynamically 
unstable with the insertion of an additional planet\footnote{In fact,
the long-term stability of the present Solar system is not guaranteed \citep{lasgas2009,davetal2013}.}.
Mounting discoveries of packed extrasolar systems 
\citep[e.g.][]{lisetal2011,lisetal2013,swietal2013,cabetal2014,masuda2014} 
suggest that the Solar system is not unusual, particularly as detection sensitivities improve and formerly 
inaccessible discovery space is becoming available for exploration.

The concept of {\it packing} has received increasing attention over the last decade.
Planet formation often appears to be efficient, even in the extreme 
environment of the first confirmed exoplanetary system (PSR B1257+12), discovered over two decades ago 
\citep{wolfra1992,wolszczan1994} around a supernovae remnant.  The ``packed planetary systems'' 
hypothesis \citep{barray2004,baretal2008} advocates that efficient formation processes create planets
with mutual separations that leave no room for additional companions.  
Dynamical scattering subsequent to formation represents another channel which can achieve tightly-packed systems
\citep[e.g.][]{rayetal2009a}.  The tally of packed systems continuously increases with observations that
rely on transit photometry \cite[][]{lisetal2011,fanmar2013}, Doppler radial velocity measurements
\cite[e.g.][]{lovetal2011,angetal2013} and even direct imaging \citep{maretal2010}.   Now planetary packing has 
become the framework in which to analyse
a variety of complex systems, such as those containing moons \citep{kanetal2013,payetal2013} or additional 
stars \citep{krasha2014}.  Theoretical interest in the formation and evolution of these packed systems has also been robust \citep{hanmur2012,hanmur2013,chatan2014,hanmur2014,hanetal2014}.

The frequent occurrence of packed systems in nature motivates analyses of their ultimate fates.  
Although we do not yet have a confirmed detection of a planet orbiting a 
WD within a few thousand au\footnote{Theoretical models constrain the mass of 
object WD 0806-661\,b, which orbits a WD at a distance of approximately 2500 au, 
to the planetary mass regime \citep{luhetal2011}.} 
\citep{muletal2008,hogetal2009,debetal2011,faeetal2011,steetal2011,fuletal2014},
we know that between 25\%-50\% of all white dwarfs (WDs), the end product of stellar evolution for over 90\% 
of all Milky Way stars, contain remnant planetary systems. This remarkable statistic follows from 
observations of abundant photospheric metal pollution in WD atmospheres \citep{zucetal2003,zucetal2010,koeetal2014}.
These metals must be accreted from circumstellar matter, which is detected in a number of cases in the form of dusty 
\citep{zucbec1987,becetal2005,kiletal2005,reaetal2005,faretal2009} and gaseous discs  
\citep{ganetal2006,ganetal2007,gansicke2011,faretal2012,meletal2012}.  High levels
of variability in these discs \citep[e.g.][]{ganetal2008,wiletal2014,xujur2014} on timescales of years highlight 
the current dynamical activity occuring in these systems.

The debris discs, with a typical radial extent of about 1 Solar radius, cannot have existed during the main sequence or
giant branch phases of stellar evolution.  These discs are currently thought to originate from asteroids 
that are perturbed by planets onto
highly-eccentric orbits \citep{bonetal2011,debetal2012,frehan2014}, whereupon the asteroids disrupt close 
to the WD \citep{graetal1990,jura2003,debetal2012,beasok2013,veretal2014a}.  The disc then accretes onto the 
WD \citep{bocraf2011,rafikov2011a,rafikov2011b,metetal2012,rafgar2012,wyaetal2014}, creating signatures
of metal pollution.  Pollution in WD systems without observed orbiting discs still somehow must arise from 
rocky planetesimals \citep{beretal2014}.

The presence of planets orbiting WDs on tight orbits would help explain asteroid delivery,
but also raise the possibility of harbouring life by residing in a WD habitable
zone \citep{monteiro2010,agol2011,barhel2013}.  
\cite{fosetal2012} has demonstrated that
photosynthetic processes associated with complex life can be sustained on these planets.  
Further,
because habitable WD planets would reside within 0.1 au, these planets are in principle 
more easily detectable
by transit \citep{agol2011} and by polarized or reflected light \citep{fosetal2012}.  
Biomarkers in the atmospheres of habitable planets transiting WDs might be observable with the James 
Web Space Telescope \citep{loemao2013}.  Indeed, JWST is capable of detecting objects with masses three 
orders of magnitude less massive than the Moon \citep{linloe2014}. However, so far, transit searches have been unsuccessful \citep{faeetal2011,fuletal2014},
and one theoretical prediction based on a simulation suite of three Jovian-mass planets
suggests that less than 1 per cent of WDs host close-in Jupiters \citep[see Section 4.3 of][]{musetal2014}.

Motivated by both the robust signatures of remnant planetary systems and the possibility
of detecting close-in planets, we here explore the
link with packed planetary systems on the main sequence.  First, however, we will 
review how long-term instability manifests itself in multi-planet systems, and comment on tidal effects between close-in planets and WDs.

\subsection{Multi-planet instabilities}

Before considering how multiple planets become unstable in post-main-sequence systems,
we should understand instability on the main sequence \citep{davetal2013}.  During
a star's main sequence lifetime, the star's mass and radius change negligibly.
These changes translate into little dynamical excitation amongst orbiting bodies;
\cite{verwya2012} point out that the Solar system planets would increase their semimajor
axes by at most about 0.055 per cent.  Due to such small changes, the vast majority of
all main sequence exoplanet studies treat the stellar mass and radius as fixed.
However, after the star leaves the main sequence, then the mass loss becomes extreme
-- causing orbital expansions by tens to several hundred percent, or outright ejection \citep{veretal2011,adaetal2013} -- 
and the stellar radius increases by several orders of magnitude.  Consequently, destabilization
of planetary systems becomes more likely.  In the Solar system, Mercury and Venus will
be swallowed by the Sun, and Earth might too be destroyed \citep{schcon2008}.

Here, we do not account for the gravitational effect of any 
residual smaller masses, such as planetesimals, asteroids or comets, on the planets:  we 
treat our initial systems as already dynamically settled from formation.  In reality, 
complex interactions between planetesimals
and planets likely play important dynamical roles, primarily at the earliest main sequence ages;
their incorporation enormously increases the available parameter space and computational cost 
of simulations \citep{tsietal2005,rayetal2009b,rayetal2010,levetal2011}.

Stability boundaries are well-defined for two-planet systems.
If the distance between two planets exceeds the Hill stability boundary, than those planetary orbits
will remain ordered (non-crossing) forever \citep[for a review, see][]{georgakarakos2008},
but not necessarily stable: the inner planet could collide with the star or the outer planet
could escape.  If that distance instead exceeds the Lagrange
stability boundary, then the planets will forever remain bounded and ordered.  The Hill
stability boundary can be expressed in Jacobi coordinates for arbitrary eccentricities and 
inclinations \citep{donnison2011} and is entirely analytical, except for a usually-negligible
truncation in the expression of the energy \citep[Fig. 19 of][]{veretal2013a}
\footnote{Perhaps related is \cite{marzari2014}'s frequency map analysis finding that the dependence 
of the Hill boundary on planet mass has not yet fully been accounted for.}.  Unfortunately,
no such analytical formulation exists for Lagrange stability.  However, empirical estimates
\citep{bargre2006,vermus2013} help establish this boundary.  Also, two planets which are not Hill
stable may still be bounded and ordered over long timescales.  This behaviour is especially
connected to the overlap of mean motion resonances 
\citep{chirikov1979,wisdom1980,mardling2008,funetal2010,muswya2012,
decetal2013,giuetal2013,bodqui2014}.

For systems with more than two planets, stability estimates become decidedly more imprecise.
No analytical criterion exists like for the two-planet Hill stability case.  Consequently,
several authors \citep{chaetal1996,fabqui2007,zhoetal2007,chaetal2008,smilis2009,quillen2011} have attempted to 
fit empirical instability timescales to systems of three or more planets from numerical simulations.
Because these estimates are primarily based on simulations with equal-mass planets,
here we also adopt equal-mass planets in order to utilize these estimates.  The rich dynamics in
systems with a (more realistic) set of unequal-mass planets attests to its greater number of 
degrees of freedom \citep{verarm2005,verarm2006,forras2008,jurtre2008,rayetal2011,rayetal2012,matetal2013}.

The dynamical complexities increase in post-main sequence systems, for which 
relatively few multi-planet studies have been performed.  All such studies consider 
gravitational scattering which results from instability
due to mass loss. \cite{dunlis1998}
investigated the future evolution of collections of planets in artificial Solar systems for various planet-Sun
mass scalings.  \cite{debsig2002} evolved systems with two and three equal-mass planets over 1000 orbits 
around a progenitor $1M_{\odot}$
star that lost half of its mass in a uniform fashion.  \cite{veretal2013a} evolved
two-planet systems with equal planetary masses from the endpoint of formation to several Gyr into the WD phase
for progenitor stellar masses of $3-8 M_{\odot}$;
this work was recently extended with simulations of equal-mass planets in three-planet systems \citep{musetal2014}.
\cite{voyetal2013} instead considered a more analytical approach, tracking the 
regular and chaotic trajectories in phase space that results from the inclusion of mass
loss in multi-planet systems.

\subsection{Tidal effects}

Planets scattered close to their parent stars might change their orbits due to star-planet
tides.  Main sequence and giant branch studies dominate investigations of star-planet tidal dynamics;  
few references to WD-planet interactions exist.  \cite{norspi2013} pose in their Section 4 
that an Earth-mass planet driven to within half of a Solar radius (about 0.002 au) of an WD 
{\it might} be circularized due to tidal friction, but do not elaborate further on this process. 
They do surmise that the resulting dissipation of heat would render the planet uninhabitable, 
even if detectable. \cite{musetal2014} suggest that the dynamics of WD-planet tidal interactions 
would be nearly equivalent to that of main sequence stars, and hence feature circularization, 
because the dominant forces arise from tidal deformation of the planet and not the star.

If we make that assumption here, then all of the uncertainties which plague tidal theory 
on the main sequence carry over to WD systems.  This problem is compounded by the 
newfound inoperability or extremely limited scope \citep{efrmak2013} of classic tidal theories 
such as the constant geometric lag model \citep{goldreich1966,murder1999} and the constant time 
lag model \citep{mignard1979,hut1981}.  Consequently, a process such as tidal circularization 
of a terrestrial planet orbiting a WD must be taken on a case-by-case basis, one that is highly 
dependent on the rheology of the planet (but independent of the WD).  The circularization 
timescale can easily vary by orders of magnitude depending on this rheology \citep{henhur2014}.
Giant planets also suffer from this uncertainty, as they may contain solid cores, necessitating
the combined modeling of fluid mechanics and solid mechanics.  As observed by \cite{ogilvie2014},
Earth's tidal dissipation is dominated by a fluid layer that represents just 0.023 per cent
of the entire planet's mass. 

Because our planets here represent point masses, we make no attempt to invent interior structures, nor 
compute maximum tidal reaches nor circularization timescales.  Nevertheless, we emphasize throughout
the manuscript that we neglect these processes, which may strongly affect planets that achieve
pericentres at orbital separations from the WD that are realistically within the reach of observational
detection.

We now proceed to setup, execute and analyse our simulations in Section \ref{simsection},
discuss the results in Section 3 and conclude in Section 4.

\section{Simulations} \label{simsection}

Our simulations incorporated self-consistently the evolution of both the host star
and planetary orbits to model the progression of the planetary system throughout
the main sequence, giant branch and WD phases.  We performed integrations over the age
of the Universe, which is approximately 14 Gyr.

\subsection{Numerical code}

We used an integrator with features from the simulators used in \cite{veretal2013a}, \cite{musetal2014} and 
\cite{veretal2014b}.  The integrator contains a Bulirsch-Stoer algorithm, originally from \cite{chambers1999}, 
that is modified with stellar mass and radius interpolation between substeps.  The stellar mass and radius
profiles are computed with the {\it SSE} code \citep{huretal2000}, assuming a Reimers mass loss prescription 
with coefficient 0.5 during early giant branch phases and a superwind prescription from \cite{vaswoo1993}
along the asymptotic giant branch phase.  We assumed that the mass loss is isotropic, an excellent assumption
for the planet-star separations considered in this paper \citep{veretal2013b}.
When the star becomes a WD, the resulting (constant) radius is replaced with the WD's Roche radius assuming
the planets all have a typical terrestrial planet density of 5.5 g/cm$^3$ (see Sect. 2.2.2).  
When a planet collides with the star, the star is modelled
to absorb the planet and increase mass accordingly, affecting the subsequent dynamics of the system.
Given the long duration of our simulations, we outputted data every 1 Myr, but importantly continuously 
tracked the minimum pericentre that every surviving planet ever achieved.
We, conservatively, adopted a Bulirsch-Stoer tolerance parameter of $10^{-13}$.
Consequently, in every case, we have found that angular momentum was conserved to within $10^{-6}$; 
energy is not conserved in mass-losing systems.

\subsection{Initial conditions}

\subsubsection{Stellar mass}

Our choice of initial stellar masses in this work ($1.5, 2.0$ and $2.5M_{\odot}$) is an improvement 
over previous studies \citep{veretal2013a,musetal2014}.  The average mass of a WD in the present-day WD 
population is $0.60-0.65M_{\odot}$
\citep{lieetal2005,faletal2010,treetal2013}, which corresponds to 
an A-type main-sequence progenitor star with a mass around $2M_{\odot}$
\citep{catetal2008,kaletal2008,casetal2009}.
Figure 1 of \cite{koeetal2014} illustrates that by considering
the stellar mass range of $1.5-2.5M_{\odot}$, we sample the progenitor
mass range of nearly all known polluted WDs.  This mass range corresponds to
main sequence lifetimes between about 600 Myr and 3 Gyr.  Consequently, 
the majority of the integration time was during the WD phase of stellar evolution.
The final WD masses corresponding to these initial main sequence masses
are $0.576$, $0.637$ and $0.690M_{\odot}$; the vast majority of the mass loss occurred on
the asymptotic giant branch phase, as during the giant branch phase only 
$0.061$, $0.002$ and $0.001M_{\odot}$ was shed.

\subsubsection{Planetary mass and orbits}

We adopted equal, Earth-mass planets ($3.0035 \times 10^{-6} M_{\odot}$) in our main
suite of simulations; later we perform a few additional full-lifetime simulations with Jupiter-mass
planets ($9.5460 \times 10^{-4} M_{\odot}$) in order to showcase the differences. 
We focus on terrestrial planets because (i) at the separations probed so far, low-mass planets appear to 
be much more common than giant planets \citep{winfab2015}, (ii) observational evidence for their packing is greater, 
and (iii) there already exist full-lifetime estimates for the critical survival separations 
throughout main sequence evolution for four or more terrestrial planets \citep{smilis2009}.

Observationally, we are not yet sensitive to these planets\footnote{See the Exoplanet
Data Explorer at http://exoplanets.org.}; terrestrial planets
in regions beyond 1 au, such as Mars analogues, remain out-of-reach.
Because super-Earths can form around Solar-type stars out to 5 au
\citep{elietal2013}, formation of Earth-like planets around more
massive stars should take place {\it beyond} 5 au.  In fact, the snow
or ice line -- a concept often used to demarcate the formation of terrestrial
from giant planets -- changes distance with time, but can reach
a distance of up to 10 au after $10^5$ yr for the stellar masses we considered
\citep{kenken2008}.  For terrestrial planets which emerge from their birth disc,
\cite{kokida1998} show that separations of 10 mutual Hill radii
are typical.  In order to estimate our mutual Hill radii, we 
utilised the definition from equation (4) of \cite{smilis2009}, which reads

\begin{eqnarray}
a_{i+1} &=& a_i 
\left[1 + \frac{\beta}{2}\left( \frac{m_i + m_{i+1}}{3 \left(m_{\star} + \sum_{k=1}^{i-1}m_{k} \right)}  \right)^{1/3}\right]
\nonumber
\\
&\times&
\left[1 - \frac{\beta}{2}\left( \frac{m_i + m_{i+1}}{3 \left(m_{\star} + \sum_{k=1}^{i-1}m_{k} \right)}  \right)^{1/3}\right]^{-1}
\label{theequation}
\end{eqnarray}

\noindent{}where $m$ and $a$ refer to mass and initial semimajor axis, subscripts refer to planets in order of increasing distance, and $\beta$ is the number of mutual Hill radii.  Four Earths orbiting a Solar-mass star with $a_1 = 1$ au, $\beta = 4$ corresponds to $(a_2, a_3, a_4 = 1.05, 1.11, 1.16)$ au whereas $\beta = 10$ instead corresponds to $(a_2, a_3, a_4 = 1.13, 1.29, 1.46)$ au.  If the planets were instead Jovian, then $\beta = 10$ would correspond to $(a_2, a_3, a_4 = 2.51, 6.29, 15.8)$ au.

The process of unpacking is largely scale-invariant, i.e. independent of our choice
of $a_1$.  However, there do exist physical constraints. \cite{johetal2012} and \cite{petetal2014} show 
an increased incidence of collisions with the central star as $a_1$ is decreased, a phenomenon 
which can be explained through the Safronov
number \citep{safzvj1969}.  This number represents the square of the ratio of orbital speed to surface escape speed.
As this ratio increases, the frequency of collisions increases.

Another constraint includes star-planet tides
along the post-main-sequence phases, as a star's radius will expand significantly and might swallow the
planet.  For the stellar mass range we consider ($1.5-2.5M_{\odot}$), a 
terrestrial planet with a circular orbit and main sequence semimajor axis of at least about 3 au will
survive engulfment from the expanding giant star \citep[Figure 7 of][]{musvil2012}.
This bound remains robust even for different mass loss prescriptions
\citep{adablo2013,viletal2014}.  Another constraint is the age of the Universe, which limits
the total number of orbits which could be completed as a function of $a_1$,
affecting Lagrange-type instability \citep[e.g.][]{vermus2013}. However, as the number of
orbits increase, computations take longer and may accumulate error in angular
momentum conservation.

For these reasons, we choose $a_1 = 5$ au for most of our simulations;
$\beta = 10$ then gives $(a_2, a_3, a_4 = 5.67, 6.44, 7.30)$ au.

\begin{table}
 \centering
  \caption{Summary of simulations, four per row.  Each simulation contains either Earth-mass planets
or Jupiter-mass planets, separated by a mutual Hill separation indicated by $\beta$ through equation (\ref{theequation}).
If $\beta$ is too large, instability will never occur.  If $\beta$ is too small, the first instability will occur
quickly (i.e., on the main sequence; MS).
}
  \begin{tabular}{@{}ccccccc@{}}
  \hline
  \hline
 \multicolumn{6}{c}{Terrestrial planets}   \\
  \hline
  \hline
Row &  \# of & progenitor &  $\beta$   & $a_{1}$ &  \% unpacked \\
 \#   &  planets & mass ($M_{\odot}$)  &   & (au) & on post-MS \\ 
 \hline
1&  4  & 1.5 & 10  & 5   & 75     \\
2&  10  & 1.5 & 10  & 5  & 100    \\
3&  4  & 2.0 & 10  & 5   & 100    \\
4&  10  & 2.0 & 10  & 5  & 100    \\
5&  4  & 2.5 & 10  & 5   & 100    \\
6&  10  & 2.5 & 10  & 5  & 100    \\
7&  4  & 1.5 & 10  & 10  & 25     \\
8&  10  & 1.5 & 10  & 10 & 100    \\
9&  4  & 1.5 & 12  & 5   & 25     \\
10&  4  & 1.5 & 15  & 5   & 0      \\
11&  4  & 2.0 & 12  & 5   & 100    \\
12&  4  & 2.0 & 15  & 5   & 25     \\
13&  4  & 2.5 & 12  & 5   & 100    \\
14&  4  & 2.5 & 15  & 5   & 0      \\
  \hline
  \hline
 \multicolumn{6}{c}{Giant planets}   \\
  \hline
  \hline
15 &  4  & 1.5 & 8  & 5   & 0     \\
16 &  4  & 2.0 & 8  & 5   & 0     \\
17 &  4  & 2.5 & 8  & 5   & 0     \\
18 &  4  & 1.5 & 6  & 5   & 100     \\
19 &  4  & 2.0 & 6  & 5   & 100     \\
20 &  4  & 2.5 & 6  & 5   & 100     \\
21 &  4  & 1.5 & 4  & 5   & 0     \\
22 &  4  & 2.0 & 4  & 5   & 0     \\
23 &  4  & 2.5 & 4  & 5   & 0     \\
  \hline
  \hline
\end{tabular}
\end{table}

We do not include Galactic tides in our simulations, as planets
in the Solar neighbourhood are negligibly perturbed by these tides
for separations up to thousands of au \citep{kaietal2013,vereva2013a,vereva2013b},
and adiabatic mass loss will increase the outermost planet's initial semimajor axis 
[$a_{10} < 20$ au] by a factor of just a few.
The outcomes of our simulations justify this choice {\it ex post facto}.  
We also do not
include flybys from passing stars, for both computational reasons and so that
we can focus on the unpacking process.  In reality, we should expect, over the 
age of the Universe, one Galactic field star to come within a few hundred 
au of another \citep{zaktre2004,vermoe2012}.  However, post-main-sequence
mass loss occurs for just a small fraction of this time.  Therefore, during
post-main-sequence mass loss, a stellar flyby would not be expected to penetrate
to within $10^4$ au in the Solar neighbourhood \citep{veretal2014c}.

Our simulations also do not include the effect of general relativity (GR).  During
the course of each orbit, GR can displace a highly eccentric planet 
from its predicted pericentre location around a typical
$0.6 M_{\odot}$ WD by up to 2.65 km \citep{veras2014}.  This value
is independent of the planet's semimajor axis or pericentre.  Other
changes to the orbit during one complete period are similarly minor.  However,
over many orbits, the pericentre may precess noticeably; the extent of this precession 
is in fact dependent on the semimajor axis and eccentricity.  Equation 19 of \cite{veretal2014a}
demonstrate that any highly-eccentric object approaching the Roche radius of the WD 
can experience precession on the order of a few degrees within $10^5$ yr.
Consequently, any secular behaviour which is strongly dependent on the argument of pericentre,
such as the Lidov-Kozai mechanism, could be affected by GR.

Unlike \cite{smilis2009}, we initialized our planets on slightly non-coplanar orbits.  We drew inclination 
values from random uniform distributions with ranges of $[-0.5^{\circ},0.5^{\circ}]$ in order
to avoid an artificially high number of collisions, given that real systems are not exactly coplanar
but tend to harbour planetary orbits which are mutually inclined on the order of degrees$^{\ 4}$.  
We performed a similar 
random selection for all orbital angles (mean anomaly, argument of pericentre, longitude of ascending node) 
over their entire ranges.  We set the planets on initially circular orbits for simplicity
and because \cite{jurtre2008}, among others, suggest that qualitative scattering outcomes are independent
of initial eccentricity.

%%%%%%%%%%%%%%%%% Figure
\begin{figure*}
\centerline{
\psfig{figure=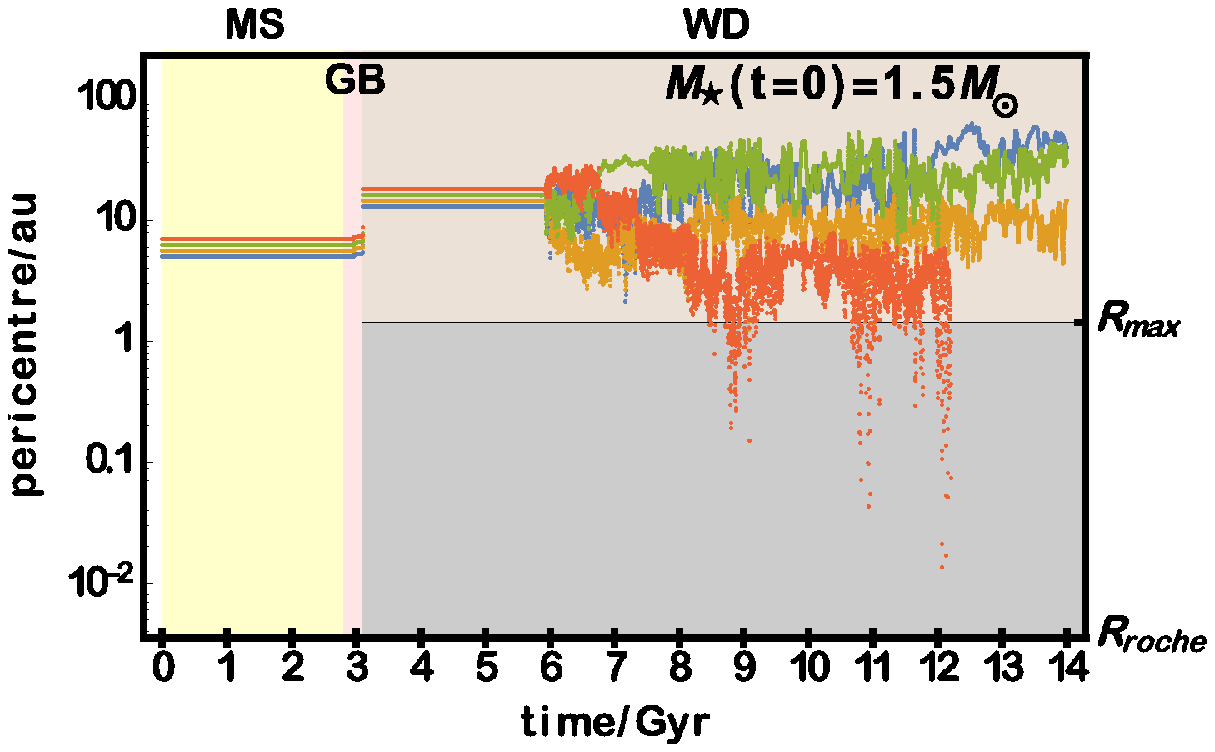,width=9.2cm} 
\psfig{figure=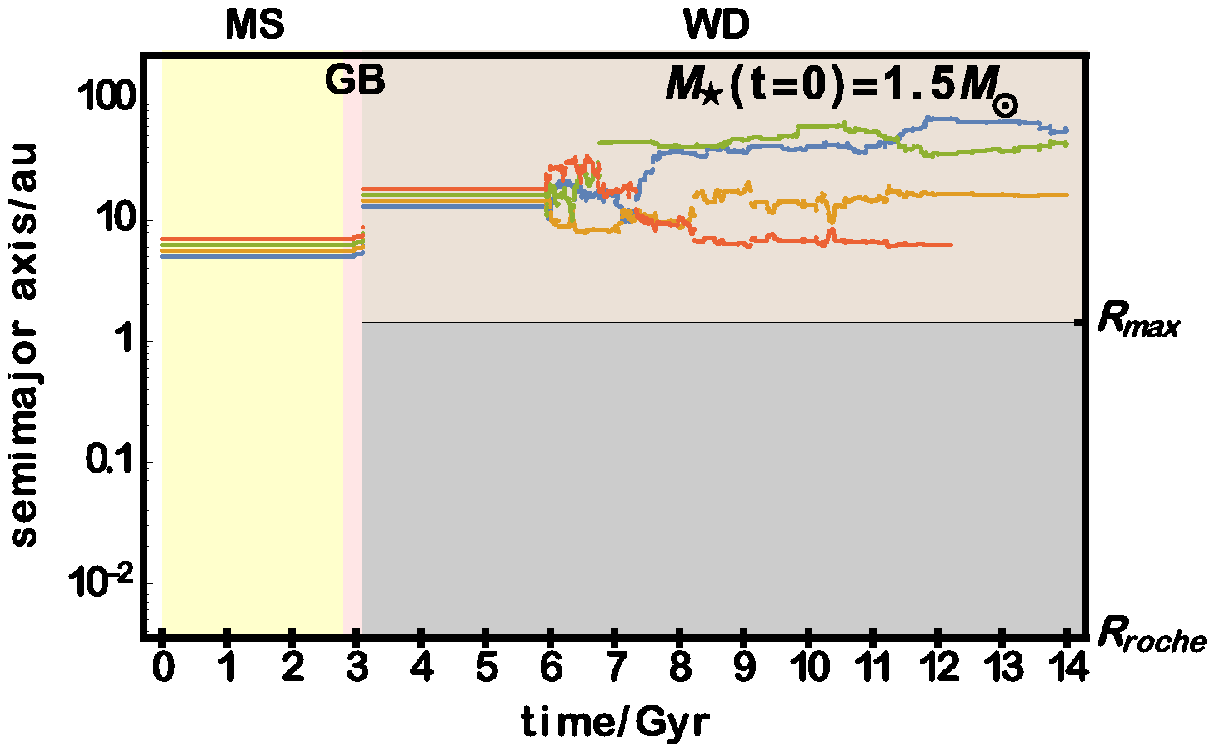,width=9.2cm}
}
\centerline{
\psfig{figure=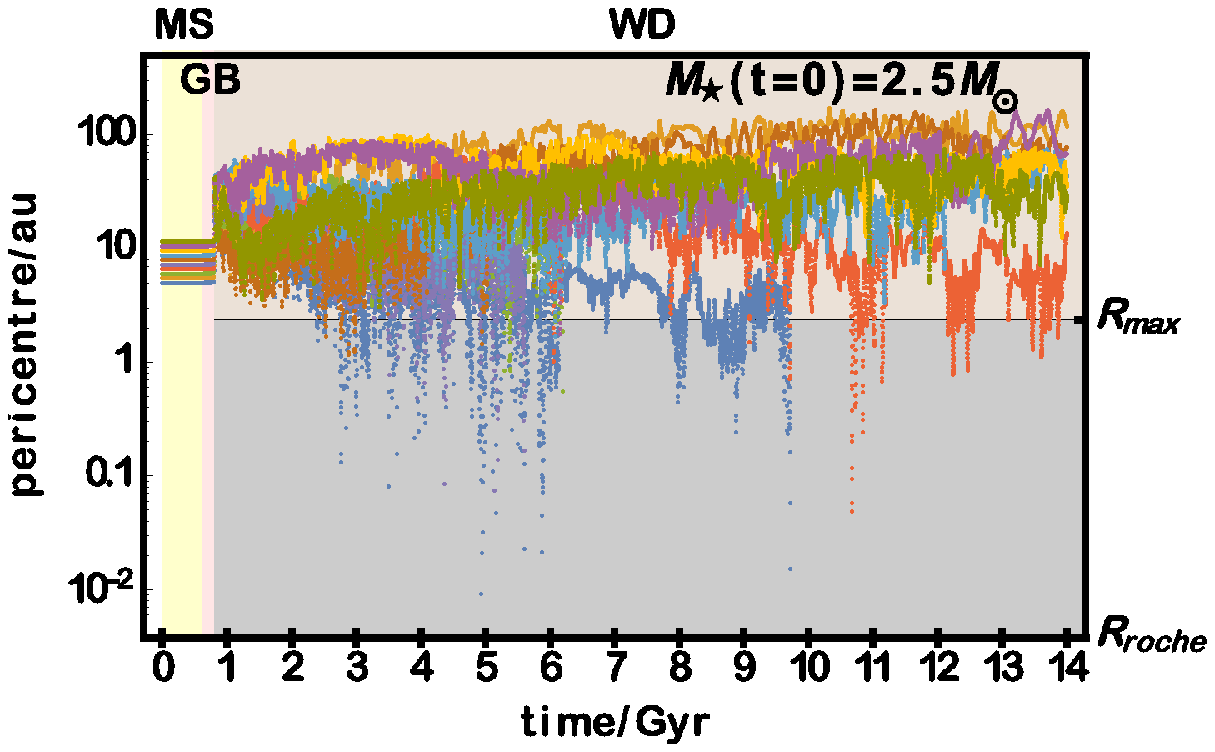,width=9.2cm} 
\psfig{figure=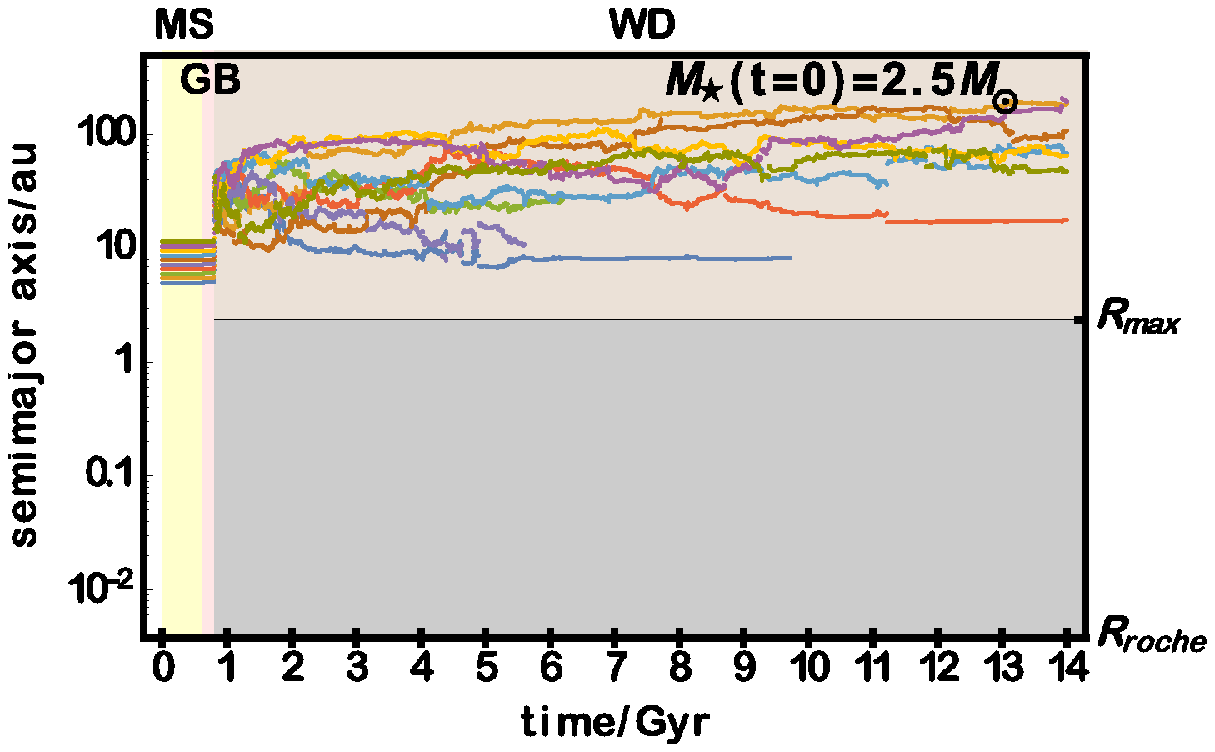,width=9.2cm}
}
\centerline{}
\caption{
Two examples of post-main-sequence planetary unpacking
with terrestrial planets.
The top and bottom panels each correspond to one simulation
from rows \#1 and \#6 of Table 1, respectively. 
The background colours refer to the main sequence (MS; yellow),
giant branch (GB; pink) and white dwarf (WD; gray) phases
of stellar evolution.
In the top panels, the 4 planets remain packed for the entire MS 
and GB phases, and for about 3 Gyr of WD 
evolution. In the bottom panels, the 
10 planets become unpacked during the GB phases (specifically, during
the Asymptotic Giant Branch).  
In both cases, planets make sweeping incursions well within
the initial innermost semimajor axis and maximum stellar radius $R_{\rm max}$, 
and at WD cooling ages exceeding 5 Gyr.  
%The innermost
%pericentre that was acquired of the three {\it surviving} planets in the
%top panel were $1.00$ au (blue, at 10.8 Gyr), $2.03$ au 
%(yellow, at 7.18 Gyr) and $5.21$ au (green, at 11.4 Gyr);
In the top left panel, the initially outermost planet (red) spent Gyrs of WD
cooling time at close pericentres before being engulfed into the
WD at 12.2 Gyr. The bottom-left panel features a particularly
intrusive incursion: the dark blue planet achieves a pericentre
of 0.046 au -- close to a typical WD habitable zone 
and beyond the WD Roche radius $R_{\rm roche}$ -- at 10.68 Gyr, and survives.
Depending on the internal properties of the planet, it might be subsequently
tidally circularized - a process ignored in this work.
The semimajor axis evolution showcase the complex energy exchanges triggered
in these systems after the first instance of
instability, and demonstrate that close pericentre passages
are due to variations in orbital eccentricity. 
}
\label{fisfig1}
\end{figure*}
%%%%%%%%%%%%%%%%% Figure

%%%%%%%%%%%%%%%%% Figure
\begin{figure*}
\centerline{
\psfig{figure=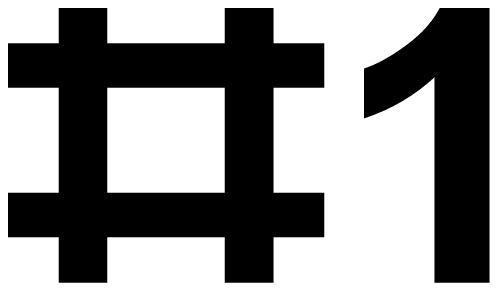,width=1.00cm} 
\psfig{figure=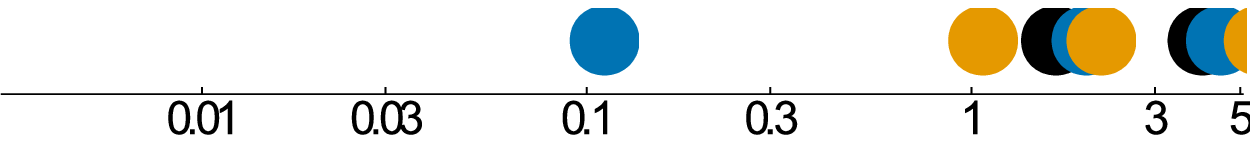,width=8.00cm} 
\psfig{figure=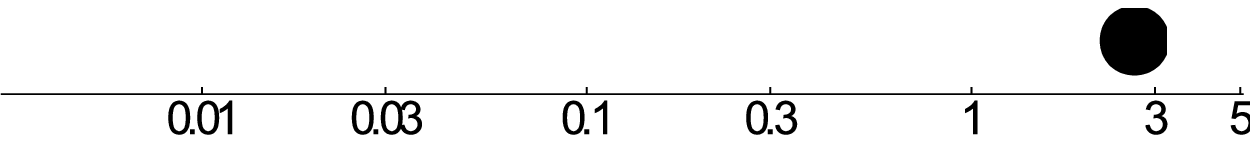,width=8.00cm} 
\psfig{figure=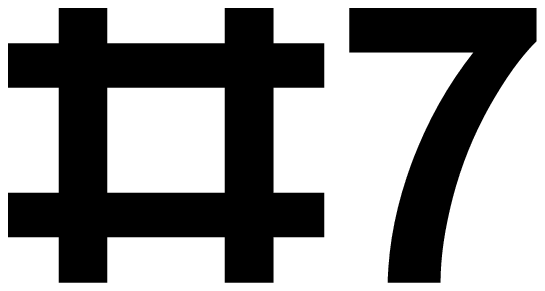,width=1.00cm} 
}
\vspace{0.3cm}
\centerline{
\psfig{figure=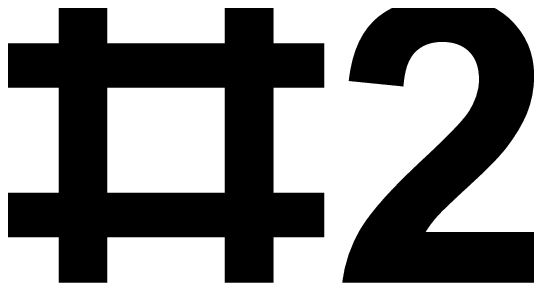,width=1.00cm} 
\psfig{figure=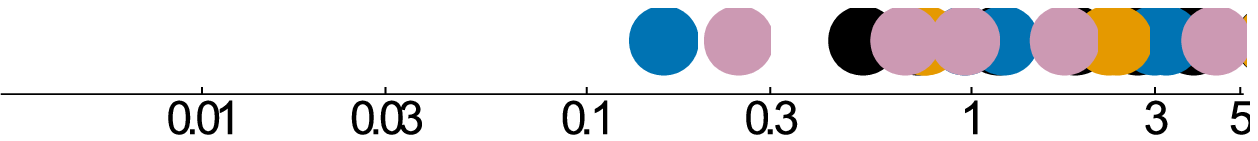,width=8.00cm} 
\psfig{figure=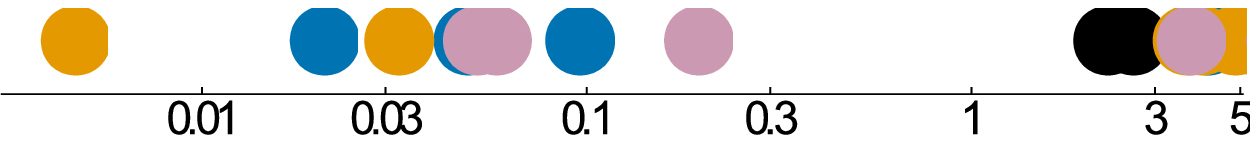,width=8.00cm} 
\psfig{figure=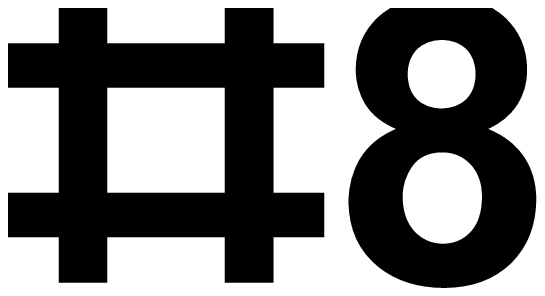,width=1.00cm} 
}
\vspace{0.3cm}
\centerline{
\psfig{figure=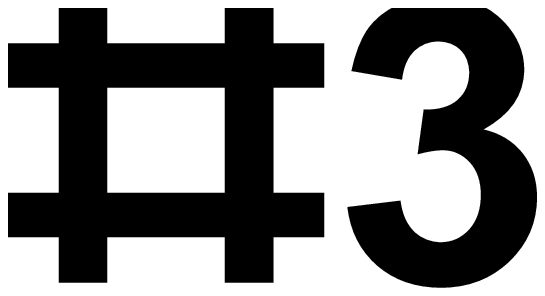,width=1.00cm} 
\psfig{figure=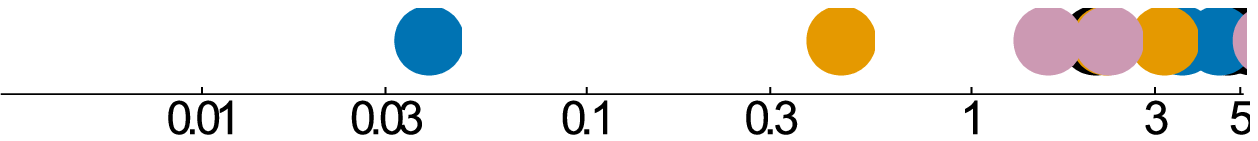,width=8.00cm} 
\psfig{figure=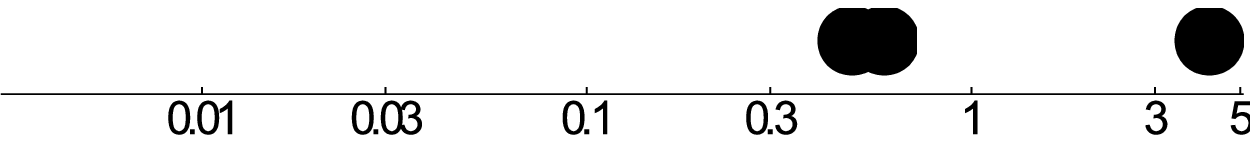,width=8.00cm}
\psfig{figure=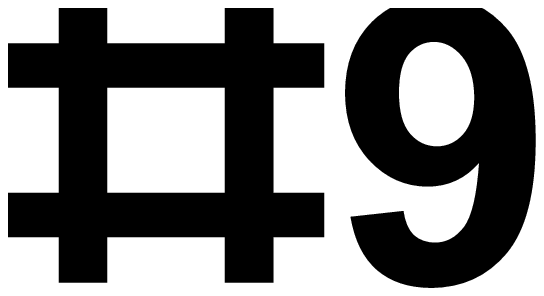,width=1.00cm} 
}
\vspace{0.3cm}
\centerline{
\psfig{figure=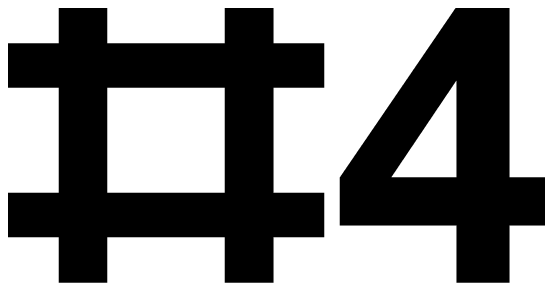,width=1.00cm} 
\psfig{figure=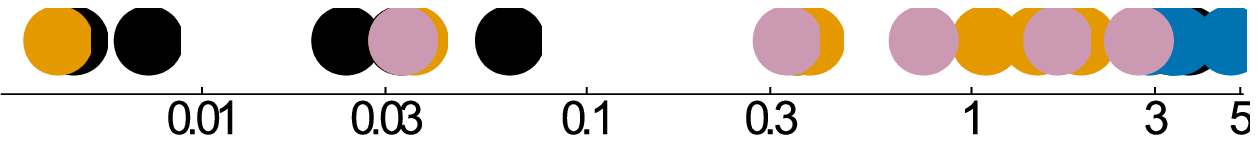,width=8.00cm} 
\psfig{figure=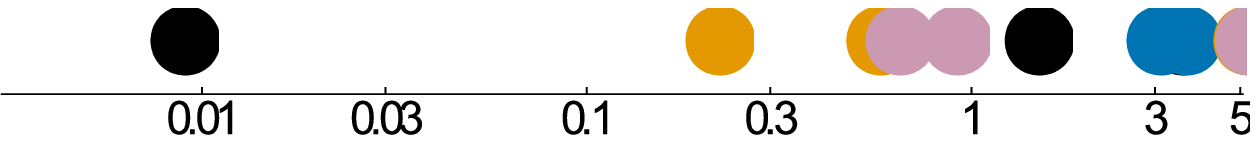,width=8.00cm} 
\psfig{figure=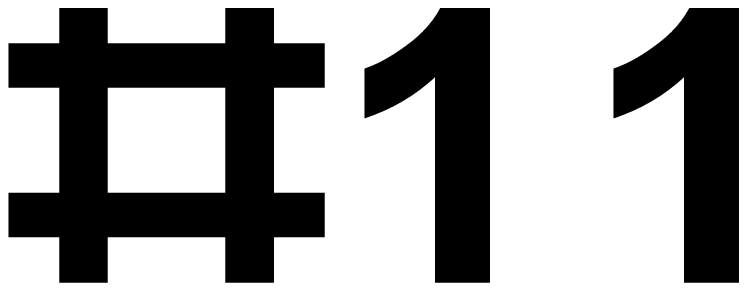,width=1.00cm} 
}
\vspace{0.3cm}
\centerline{
\psfig{figure=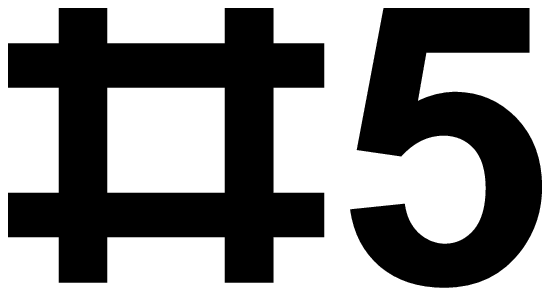,width=1.00cm} 
\psfig{figure=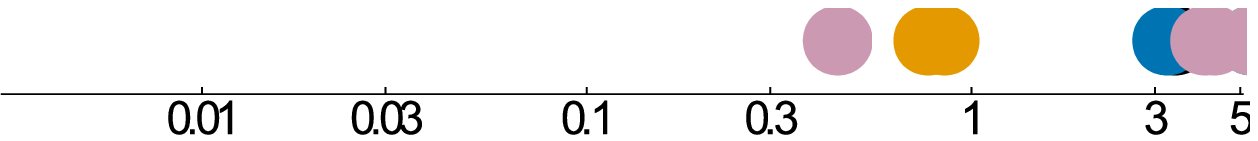,width=8.00cm} 
\psfig{figure=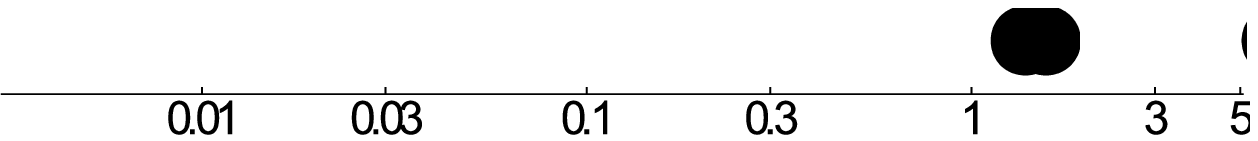,width=8.00cm} 
\psfig{figure=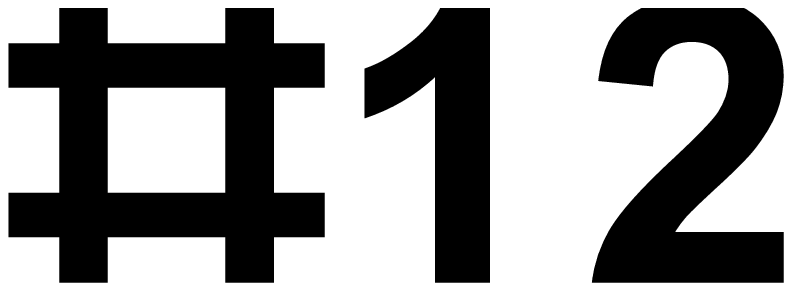,width=1.00cm} 
}
\vspace{0.3cm} 
\centerline{
\psfig{figure=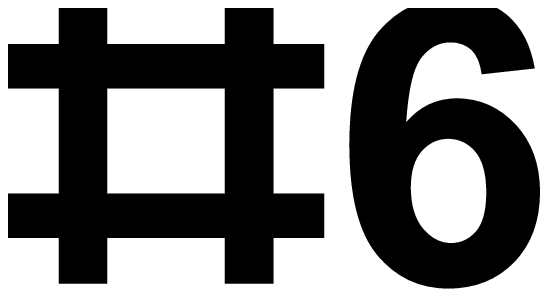,width=1.00cm} 
\psfig{figure=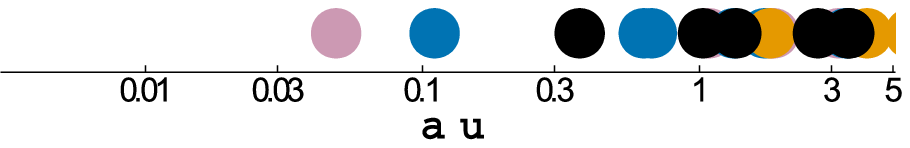,width=8.00cm} 
\psfig{figure=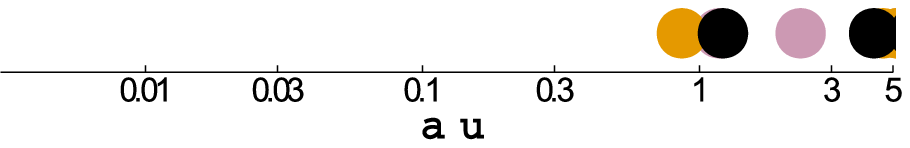,width=8.00cm} 
\psfig{figure=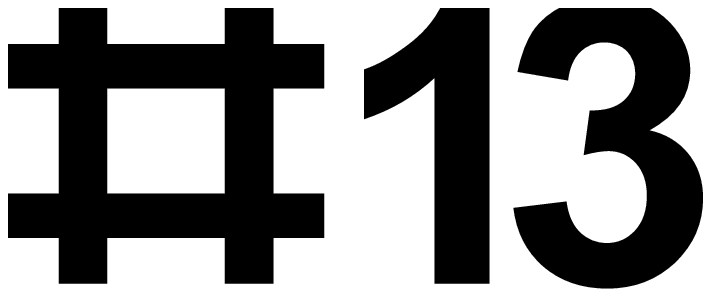,width=1.00cm} 
}
%\centerline{
%\psfig{figure=LeftColumn.eps,width=5.00cm} 
%\psfig{figure=FinalPlot.eps,width=10.17cm} 
%\psfig{figure=RightColumn.eps,width=5.0cm} 
%}
%}
\caption{
Minimum pericentres (in au) which are less than 5 au 
for systems in Table 1.  Each numberline corresponds 
to a different (labelled) row of the table (rows \#10 and \#14 did
not unpack), and 
each numberline displays all planets in all four runs of a given row;
each run is colour-coded.
Note that even for rows \#7 and \#8, where $a_1 = 10$ au, at least one planet
achieves a minimum pericentre within 5 au.  This figure 
demonstrates that an endemic feature of post-main-sequence 
unpacking is inward incursions of one or more planets to 
values which are just a fraction of their original main 
sequence semimajor axes.
}
\label{fisfig2}
\end{figure*}
%%%%%%%%%%%%%%%%% Figure

\subsection{Simulation results}

\subsubsection{Terrestrial planet suite}

Here we discuss the results of our main set of simulations, summarized in Table 1, 
and show two illustrative evolution examples in Fig. \ref{fisfig1}.  
Figure \ref{fisfig2} summarizes the results from the majority of the simulations
in the table.  We performed four simulations with slightly different initial conditions (Sect. 2.2.2) for each set of parameters, i.e. for each row
in Table 1.  Within each row, the number of planets as well as the values
of $M_{\odot}, \beta$ and $a_1$ were fixed.

The table reveals that our
choice of 10 mutual Hill radii successfully keeps the planets ordered and
packed along the main sequence, before unpacking occurs along the 
post-main-sequence phases.  Rows \#9, \#11 and \#13 indicate that increasing
$\beta$ to 12 reproduces similar behaviour.  When $\beta = 15$, the planets
are too widely separated to ever unpack, except in a single 
case (see rows \#10, \#12 and \#14).  The finding that unpacking ceases
when $12< \beta < 15$ is in line with the prediction from Fig. 11
of \cite{musetal2014} after their single-planet Hill radii is converted
into the mutual-planet Hill radii used here (originally from \citealt*{smilis2009}).
Doubling the value of $a_1$, as
in rows \#7-\#8, still produces unpacking in all four 10-planet cases,
but in just one of the 4-planet cases.  The likely reason is because these
planets have not completed as many orbits 
(see Fig. 6 of \citealt*{veretal2013a} and \citealt*{vermus2013}).  Altering 
the stellar mass does not
have a significant effect on the final unpacking statistics, but dramatically
changes the main sequence lifetimes and therefore the possible extent of the WD
cooling ages.  Nevertheless, comparing the last column in the first 6 
rows suggests that the
amount of mass lost from $1.5 M_{\odot}$ progenitor mass stars was just not high
enough to produce unpacking in every case (Row \#1)\footnote{More precisely,
in one case the mass loss failed to shift the stability boundary sufficiently 
enough to cause orbit crossing either during a giant branch or WD phase.}.

Figure \ref{fisfig1} provides two examples of the time evolution of 
individual simulations from rows \#1 and \#6.  The plots visually
demonstrate that the main-sequence lifetimes of the $1.5 M_{\odot}$
and $2.5 M_{\odot}$ stars are roughly 3 Gyr and 0.6 Gyr.  Along this
phase, the planets remain ordered and packed.

Now consider the upper panels in more detail.  The 4 planets here 
remain in this effectively stable state throughout both the giant
branch phases, when their semimajor axes are more than doubled, and
through nearly 3 Gyr of WD cooling.  Only after this time do the planets
unpack, crossing one another's orbits.  Subsequently, the planets' 
orbital eccentricities are excited, and some planets sweep inwards
towards the WD at distances much less than the original innermost
orbit ($a_1 = 5$~au), or even the maximum stellar radius 
($R_{\rm max} = 1.42$ au).
Eventually, the initially outermost planet enters the Roche radius
of the WD ($R_{\rm roche} = 0.0035$ au) at 12.2 Gyr, 
corresponding to a cooling age of about 9 Gyr.

Contrastingly, in the bottom panels, the initially packed system of 10 planets becomes 
unpacked during the giant branch phases, and specifically during the asymptotic giant
branch phase\footnote{None of our simulations were destabilized along the Red Giant
Branch phases.}.  With 10 planets and a high maximum amount of total mass lost, 
the stability boundary is more easily critically shifted than in the 4-planet
case with a smaller maximum amount of mass lost.  During the entire WD phase, lasting over 13 Gyr, 
the inner system is swept through by planets on eccentric orbits.  Three of these
planets hit the Roche radius of the WD, at 5.61 Gyr, 6.19 Gyr and 9.72 Gyr,
at which point the planets are absorbed into the WD and increase its mass.
One of the planets (red) which survives for the age of the Universe drifts within
0.05 au of the WD at 10.68 Gyr.  A planet with such a close pericentre has a 
non-negligible probability of being detected while transiting the WD.
If the pericentre is close enough to the WD for tidal circularization to occur,
then prospects for detection increase dramatically.  As the orbit circularizes, the geometric transit probability remains the same, while the transit duration increases and the orbital period decreases \citep{barnes2007,socetal2012}; the latter effects provide the enhancement in detectability.

All instabilities manifested themselves as planet-planet collisions or planet-star
collisions; no planets were ejected.  Ejections require planets to be scattered
out to large distances from the star, and these distances were not attained with the
terrestrial planets simulated here.  Our integrator faithfully tested for ejections by 
incorporating the true Hill escape ellipsoid in the Solar neighbourhood, with
semi-axes on the order of $10^5$ au \citep{vereva2013b}.  Therefore, planets
flung out on orbits of several hundred au are retained (see, e.g. the bottom 
right panel of Fig. \ref{fisfig1}).  Further, \cite{forras2008} suggest that 
the smaller the planets, the less likely ejection is to occur.  Finally, stellar
mass loss by itself is not fast and great enough to cause escape due to both
the physical properties of the stars simulated and the planetary orbits considered
\citep{veretal2011}.  Of the instabilities across all terrestrial planet simulations, 
three quarters were engulfments into the star, and one quarter were planet-planet 
collisions.  Planet-planet collisions are a plausible avenue to generate large amounts of fresh small body debris, which could explain strong metal pollution detected in a number of old, cool WDs \citep{koeetal2011}.

We can improve our understanding of the dynamical evolution in both panels by 
focusing on the planetary semimajor axis changes throughout the star's lifetime
(right panels of Fig. \ref{fisfig1}).
Although a system's total orbital energy includes 
eccentricity-dependent and inclination-dependent
planet-planet interaction terms (see e.g. equation 2.27 of 
\citealt*{veras2007}), we can use the other terms, whose only
orbital parameters are semimajor axes, as excellent 
proxies (see Fig. 19 of \citealt*{veretal2013a}) for energy
transfers in the system.  Further, after the star has
become a WD, and mass loss ceases, energy is conserved for the remainder
of the system's existence.  In
Fig. \ref{fisfig1} the semimajor axes ``jump'' around in a vaguely sawtooth-like
manner, indicating shifts in orbital energy due to instances of strong 
interactions with the other planets.  

In the upper right panel, after an 
initial period of orbit
crossing lasting a few Gyr, the planets' orbits remain largely ordered,
except for one notable orbit crossing at just over 11 Gyr.
In no instance is the semimajor axis of a planet perturbed down to values
within a few au.  This result is expected because
when the planets' orbits are ordered and not undergoing strong interactions,
they are evolving secularly, and secular interactions do not alter semimajor axes.
Also, during scattering the apocentre must remain high in order to interact with 
the other planets.  Hence, the semimajor axis cannot be reduced by much more than
50 per cent.  Therefore, the small pericentres achieved are due 
primarily to changes in orbital eccentricity and not semimajor axis.
Nevertheless, this analysis neglects the effects from tidal circularization,
which may effectively ``grab'' a planet at its pericentre and keep 
that pericentre to within a factor of 2 while shrinking the orbital 
semimajor axis.

Regardless of semimajor axis, the minimum planetary pericentre is the value of greatest interest 
here because of (i) how close-orbit planets might potentially perturb asteroids into the
Roche radius of the WD, and (ii) the possibility of detecting these 
planets \citep{agol2011,fosetal2012}.  Therefore, we present
a collection of these minimum pericentres in Fig. \ref{fisfig2} for all simulations
that became unpacked and for all planets that have survived for the age of the Universe.
For each of 7 different numberlines (sets of initial conditions), at least
one simulation featured an incursion within 0.1 au of the star.  In one 10-planet
simulation (row \#2), 5 of the surviving planets each individually drifted 
to within 5 au of the star.  Each numberline demonstrates a wide variation of 
behaviour across the simulations, perhaps emphasising the highly chaotic nature of unpacked planetary systems.

%%%%%%%%%%%%%%%%% Figure
\begin{figure*}
\centerline{
\psfig{figure=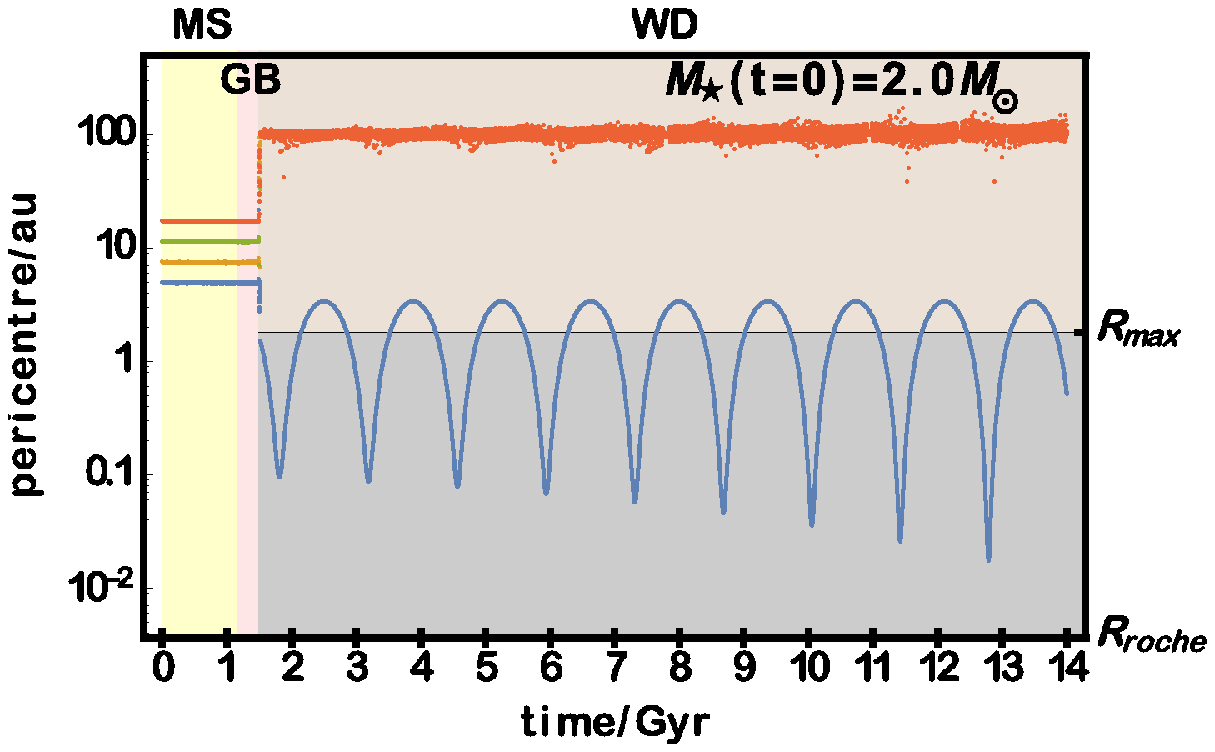,width=9.2cm} 
\psfig{figure=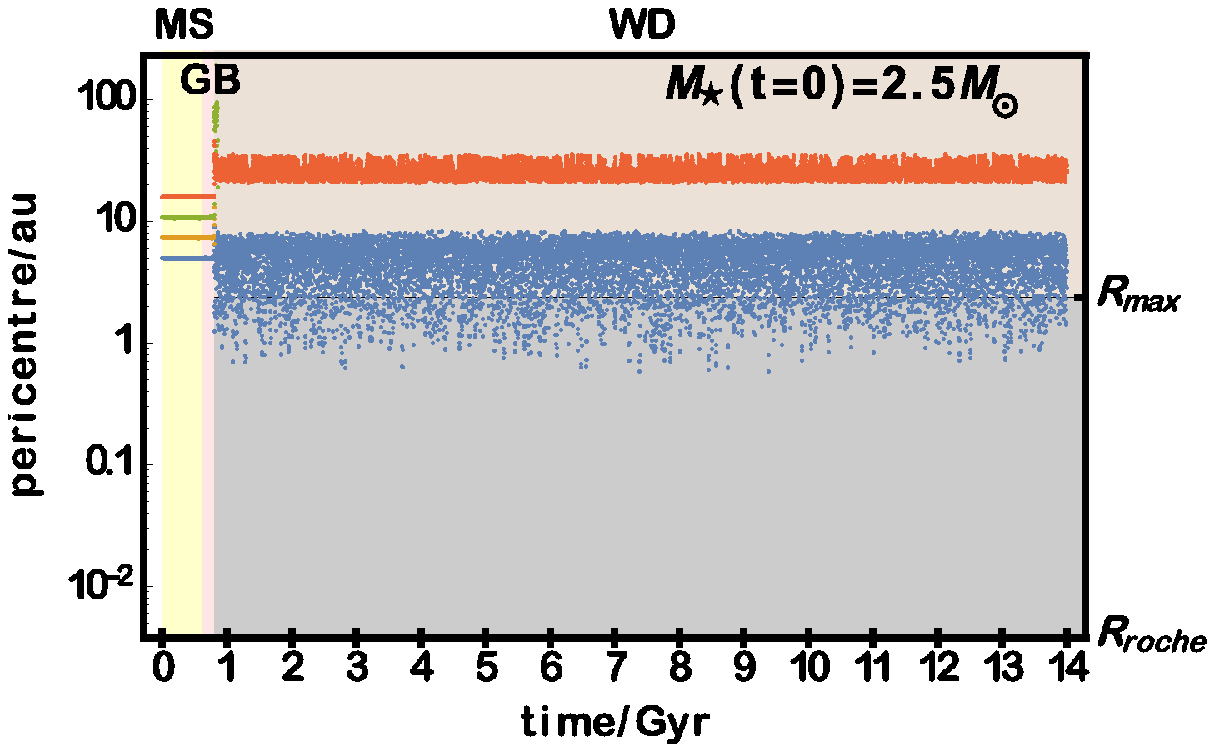,width=9.2cm}
}
\centerline{}
\caption{
Two examples of post-main-sequence planetary unpacking
with Jovian planets, from Rows \#19-\#20 of Table 1.  Both 
panels feature four planets
with initial semimajor axes between 5 au and 17.38 au (left panel)
or 5 au and 15.86 au (right panel).  In both cases, instability
occurs only after the star has become a WD and already 
shrunk its radius down from its maximum value $R_{\rm max}$ to 
approximately that of the Earth.
In the left panel, the inward incursions of the initially
innermost planet are periodic with increasing penetration.
Alternatively, in the right panel, the incursions are 
more frequent and irregular.  Neither GR nor tidal forcing
is included in these simulations, which may affect the evolution
of the simualtion in particular in the left panel.
}
\label{fisfig3}
\end{figure*}
%%%%%%%%%%%%%%%%% Figure

\subsubsection{Selected giant planet simulations}

We performed some additional simulations, this time with packed giant planets, in order
to determine how the process of unpacking changes for a higher planet/star mass ratio.  
In these simulations, we assumed a WD Roche radius based on a typical 
giant planet density of 1.0 g/cm$^3$.
There are two particularly
strong cases of packed giant planets which act as observational motivation: our solar
system, and the HR 8799 system \citep{maretal2008,maretal2010}.

We found that closely-packed giant planets which remain stable throughout the main
sequence before undergoing instability require $\beta < 8$.  
These lower values differ from the terrestrial case probably because 
the Hill sphere is no longer the relevant parameter governing the dynamics;
see equation 2 of \cite{fabqui2007} for the explicit dependence on planet mass.
The predictions for stability of post-MS systems based on this equation seem to agree 
with the results presented in the present paper for giant planets too 
(see Fig 11 of \citealt*{musetal2014}).
We also find that giant planet unpacking is more violent than terrestrial planet unpacking,
resulting in less orbit meandering, speedy dynamical settling following instability,
and escaping planets \footnote{The stronger close encounters likely decrease the accuracy
of the simulations; here, typically total angular momentum is conserved to only 
$10^{-4}-10^{-3}$.  These values indicate that future studies modeling full-lifetime
simulations containing strong interactions amongst many giant planets would benefit
from either tiny Bulirsch-Stoer tolerances ($< 10^{-13}$) -- prolonging the simulations
even further -- or by utlizing other codes.  One may combine planetary and stellar evolution
codes in the AMUSE package \citep{peletal2013}, or modify an existing planetary integrator
that emphasizes greater speed and accuracy \citep{grista2014,reispi2014}.}.

Figure \ref{fisfig3} illustrates two specific examples.  For both cases, $\beta = 6$,
and the instabilities in the system occur during the WD phase, {\it after} the stellar radius has
been reduced to approximately Earth-size.  In both cases, two planets are ejected, and one remaining
planet is perturbed onto a wide orbit.  The other planet experiences eccentric excursions 
close to the WD, and may be tidally circularized depending on the internal properties of the
planet and its orbital pericentre.  

\begin{figure}
\centerline{
\psfig{figure=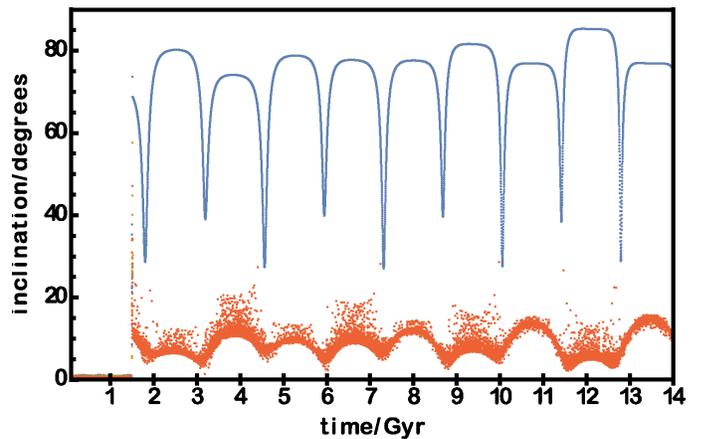,width=9.2cm} 
}
\centerline{}
\caption{
Lidov-Kozai oscillations of inclination from the left panel of 
Figure \ref{fisfig3}. The interplay between the inclination and eccentricity variations
result in close pericentre passages around the WD in this one case 
when two planets survive instability. 
}
\label{fisfig4}
\end{figure}

The manner in which the incursions 
take place can be periodic and well-defined
(left panel) or irregular and frequent (right panel).  The oscillations in the left panel
have an increasing amplitude, suggesting a way for giant planets to potentially collide with
a WD after 10 Gyr of evolution.
An event of this kind would give rise to a noticeable optical
(and probably X-ray) transient, and result in a dramatic amount of metal pollution of the
WD photosphere. Such large and periodic eccentricity variations are classic 
signatures of Lidov-Kozai oscillations (see Section 7 of \citealt*{davetal2013} for a review); 
figure \ref{fisfig4} confirms this suspicion with accompanying inclination oscillations.  
The changing amplitude may be indicative of the role
of the octupole term in Lidov-Kozai evolution \citep[see Fig. 10 of][]{naoetal2013}.
Of note is that when the orbit pericentre is at a minimum, its inclination is also at a minimum.
Then we might expect a detectable planet in this case to harbour an inclination at the critical
Lidov-Kozai value, around 40 degrees.  Because this mechanism is strongly dependent on the evolution
of the argument of pericentre, the evolution of this particular system may have been different
had we included GR in our simulations.

If tidal circularization is an active process in WD systems, then the inclinations of highly 
eccentric planets will be preserved as they are circularized.  The reason is that inclination
damping is primarily due to tides raised on the star, which are weak for WDs (as they are
for main sequence stars).  Consequently, close-in planets orbiting WDs might harbour a broad
range of inclinations.

\section{Discussion}

Our results demonstrate that planets can be perturbed onto orbits taking them close to the WD
at late ages, after an entirely quiescent existence.  This finding
is particularly important in the context of observations of {\it old} WDs with
extreme metal pollution (\citealt*{koeetal2011} and Fig. 8 of 
\citealt*{koeetal2014}) or orbiting debris discs \citep[e.g.][]{faretal2011}.
If asteroids are the progenitors of the pollution and the discs, then the asteroids must be 
flung towards the WD, and probably by a planet.

One effective way to deliver asteroids is with a planetary mean motion resonance 
\citep{debetal2012}.  Inside of a resonance, asteroid eccentricities can be
pumped high enough to eventually achieve a WD-skimming orbit.  This physical process,
when applied to a single planet and an asteroid or Kuiper belt,
produces a gradually decreasing stream of asteroids due to a depleting reservoir.  Only a massive-enough belt
would be able to sustain a high-enough delivery rate to explain accretion
in WDs which are many Gyr old.

An alternative to resonant diffusion is direct scattering.  However, a single planet
will never change its orbit, except during giant branch mass loss, due to interactions 
with a relatively massive planetesimal belt, or due to a flyby
star.  Therefore, scattering between a single planet and an asteroid belt will occur 
at young WD cooling ages \citep{bonetal2011,frehan2014}. Penetrative stellar flybys 
can scatter the planet, thereby refreshing planet-asteroid dynamics in the system, but only
for favourable geometries, and infrequently \citep{vermoe2012}.

Neither single-planet scattering, nor resonant diffusion commencing when the
WD is born, can easily explain pollution or discs which exist around WDs that are
many Gyr old.  Multi-planet systems, in concert with smaller bodies, may represent
a solution.  Two-planet systems are likely to be inadequate, because after an instance
of instability, the surviving planet will have a fixed orbit, akin to the
one-planet case.  Although this orbit can be highly eccentric and penetrate to
within the maximum radius of the star \citep{veretal2013a}, the orbit is periodic,
and will too-quickly exhaust its supply of fortuitously-located asteroids.  Three-planet
systems are more promising because after instability, two surviving planets can exist 
in ever-changing orbits for the entire WD cooling age \citep{musetal2014}, similar
to the behaviour here in the right panel of Figure \ref{fisfig3}.  Nevertheless,
if these two surviving planets are secularly evolving, they will experience predictably periodic
changes in their orbits (left panel of Figure \ref{fisfig3}).

Systems with at least four planets appear to harbour dynamics which are rich enough to 
potentially generate dynamical interactions with asteroids over long
timespans and over an extensive range of physical space.  Given the four terrestrial planets
in the Solar system, and the mounting discoveries of packed exosystems, such configurations
are perhaps common.  Indeed, over half of all stars in the Milky Way are thought to contain
at least one terrestrial planet \citep{casetal2012}, and perhaps an average of at least five
\citep{baljoh2014}.  This average may be even higher considering that \cite{baljoh2014}
focussed on M dwarfs, which have much lower masses than the stars considered here.
Further, the asteroidal material
in the remnant WD system may be widely dispersed and have a size distribution skewed
towards small bodies due to rotational fission from giant branch 
radiation \citep{veretal2014d}.  Consequently, rocky material may be available in
enough locations to be accessed at different times during WD cooling.

Our simulations also provide a potential evolutionary pathway -- a dynamical history --
for eccentric planets which could be detected transiting WDs within 0.1 au.  
If these planets fail to be tidally circularized, then they
would not remain in a secularly stable compact configuration for Gyrs.
Instead, their pericentres would drift by orders of magnitude, greatly reducing their
prospects for habitability.  Consequently, 
observerations would have to ``catch'' a system at just the right point in its secular
evolution \citep[e.g.][]{verfor2009}.  
We suggest that detectable eccentric planets
would likely be accompanied by an unseen outer companion.
The prospects for detection will increase with current and upcoming missions such
as Gaia, LSST and PLATO, which will survey $10^5-10^6$ WDs tens to thousands of times.

\section{Conclusions}

We find that the post-main-sequence peregrinations of 
terrestrial planets which were packed and 
quiescent throughout the main sequence cover a large
volume of space at both early and late WD cooling ages.  
These planets often but irregularly invade 
regions which are well-interior to their nascent orbits
and potentially contain asteroidal material that remained largely unperturbed for 
the entire main sequence and giant branch
phases of stellar evolution.  Consequently, these disturbances could represent
catalysts for dynamical interactions which lead to the formation of debris discs
and/or atmospheric pollution in old WDs.  The planets themselves, when 
traveling through the pericentres of their orbits during epochs of high eccentricity, 
should be detectable and might be tidally circularized, but 
are unlikely to harbour life.

\section*{Acknowledgments}

We thank the referee for a thorough and thoughtful review, explicit and constructive comments, 
and an expeditious response to our submission.  We also thank Alex Bowler for pointing
out a typographic error in Figure 3.
The research leading to these results has received funding from the European 
Research Council under the European Union's Seventh Framework Programme (FP/2007-2013) 
/ ERC Grant Agreement n. 320964 (WDTracer).

%%%%%%%%%%%%%%%%% Figure
%%%%%%%%%%%%%%%%% Figure

\label{lastpage}
\end{document}